\documentstyle[12pt]{article}
\newlength{\pubnumber} 

\catcode`\@=11
\@addtoreset{equation}{section}

\def\section{\@startsection{section}{1}{\z@}{3.5ex plus 1ex minus .2ex}
 {2.3ex plus .2ex}{\large\bf}}
\def\subsection{\@startsection{subsection}{2}{\z@}{2.3ex plus .2ex}
 {2.3ex plus .2ex}{\bf}}

\begin{document}

\begin{titlepage}
\samepage{
\setcounter{page}{0}
\rightline{OUTP-02-15P}
\rightline{\tt hep-th/0204080}
\rightline{April 2002}
\vfill
\begin{center}
 {\Large \bf Nonperturbative Flipped $SU(5)$ Vacua\\ in Heterotic
M--Theory\\}
\vspace{.10in}
{\Large \bf}
\vfill
\vspace{.25in}
 {\large Alon E. Faraggi$^1$\footnote{faraggi@thphys.ox.ac.uk},
        Richard Garavuso$^1$\footnote{garavuso@thphys.ox.ac.uk}
        $\,$and$\,$
        Jos\'e M. Isidro$^{1,2}$\footnote{isidro@thphys.ox.ac.uk}\\}
\vspace{.25in}
{\it $^{1}$Theoretical Physics Department, University of Oxford, Oxford
OX1 3NP, UK\\
$^2$Instituto de F\'{\i}sica Corpuscular (CSIC-UVEG)\\
Apartado de Correos 22085, 46071 Valencia, Spain\\}
\end{center}
\vfill
\begin{abstract}
  {\rm

The evidence for neutrino masses in atmospheric and solar neutrino
experiments provides further support for the embedding of
the Standard Model fermions in the chiral ${\bf 16}$ $SO(10)$ representation.
Such an embedding is afforded by the realistic free fermionic
heterotic--string models. In this paper we advance the study of
these string models toward a non--perturbative analysis
by generalizing the work of Donagi, Pantev, Ovrut and Waldram
from the case of $G=SU(2n+1)$ to $G=SU(2n)$ stable holomorphic
vector bundles on elliptically fibered Calabi--Yau manifolds
with fundamental group ${\bf Z}_2$. We demonstrate
existence of $G=SU(4)$ solutions with three generations and
$SO(10)$ observable gauge group over Hirzebruch base surface,
whereas we show that certain classes of del Pezzo base surface
do not admit such
solutions. The $SO(10)$ symmetry is broken to $SU(5)\times U(1)$
by a Wilson line. The overlap with the realistic free fermionic
heterotic--string models is discussed.

 }
\end{abstract}
\vfill
\smallskip}
\end{titlepage}

\tableofcontents
\setcounter{footnote}{0}

\def\AEF{A.E. Faraggi}
\def\NPB#1#2#3{{\it Nucl.\ Phys.}\/ {\bf B#1} (#2) #3}
\def\PLB#1#2#3{{\it Phys.\ Lett.}\/ {\bf B#1} (#2) #3}
\def\PRD#1#2#3{{\it Phys.\ Rev.}\/ {\bf D#1} (#2) #3}
\def\PRL#1#2#3{{\it Phys.\ Rev.\ Lett.}\/ {\bf #1} (#2) #3}
\def\PRT#1#2#3{{\it Phys.\ Rep.}\/ {\bf#1} (#2) #3}
\def\MODA#1#2#3{{\it Mod.\ Phys.\ Lett.}\/ {\bf A#1} (#2) #3}
\def\IJMP#1#2#3{{\it Int.\ J.\ Mod.\ Phys.}\/ {\bf A#1} (#2) #3}
\def\nuvc#1#2#3{{\it Nuovo Cimento}\/ {\bf #1A} (#2) #3}
\def\etal{{\it et al}\ }

\section{Introduction}\label{intro}

Over the past few years a profound new understanding of string theory
has emerged. In this picture the different perturbative string theories,
together with eleven dimensional supergravity, are limits of a single
underlying quantum theory \cite{mreviews}.
While the rigorous formulation of this
theory is still elusive, this development
means that we can utilize any of the perturbative string
limits to probe the features of the more fundamental structure.
In particular, we can probe those properties that pertain to the
phenomenological and cosmological features, as we observe them in
our experimental apparatus, and by using the
low energy effective field
theory parameterization. One of these properties, indicated
by the observed experimental data, is the embedding
of the Standard Model matter states in the chiral ${\bf 16}$ representation
of $SO(10)$. This embedding received in recent years additional
strong support from the evidence for neutrino masses in atmospheric
and solar neutrino experiments \cite{nexperiments}.
This embedding also yields the canonical GUT normalization
of the weak hypercharge and consequently qualitative agreement
with the measured values of $\sin^2\theta_W(M_Z)$ and $\alpha_s(M_Z)$
\cite{gcu}.

The perturbative string limit which
may preserve this $SO(10)$ embedding is the heterotic string.
It should be emphasized, however, that the heterotic string
in itself does not guarantee the preservation of the $SO(10)$
embedding and indeed many quasi--realistic models have been constructed that
do not maintain the $SO(10)$ embedding \cite{nonso10}.
A class of realistic string models that do preserve
the $SO(10)$ embedding are the free fermionic heterotic string models
\cite{rffm}.
In these three generation models the $SO(10)$ symmetry
is broken to one of its subgroups by utilizing Wilson
line symmetry breaking. A generic feature, in fact, of heterotic
string unification, with profound phenomenological and
cosmological implications \cite{fccp}, is precisely the utilization of
Wilson--line GUT symmetry breaking, rather than GUT symmetry
breaking by the Higgs mechanism. In summary,
there are two pivotal ingredients that we would like the realistic
string vacuum to possess. First, it should admit the
$SO(10)$ embedding of the Standard Model spectrum,
which is motivated by the observed experimental data.
Second, it should allow for the Wilson breaking of
the $SO(10)$ symmetry.

These two ingredients are in general not accommodated
in generic string vacua, but are afforded
by the realistic free fermionic models. The free fermionic
models are, however, constructed in the perturbative
heterotic string limit and it is therefore natural to
examine which of their structures is preserved in the
nonperturbative limit.
The nonperturbative limit of the heterotic string is conjectured
to be given by the heterotic M--theory limit, or by compactifications
of the Ho\v rava--Witten model \cite{hw} on Calabi--Yau threefolds.

We further remark that one should not expect
heterotic M--theory compactifications to compete with the
perturbative heterotic string in trying to calculate
properties of the vacuum that are more readily obtained
in the perturbative limit. Indeed, we may not even know
how or whether some of these properties are defined
in the nonperturbative limit. Thus, details of the particle
spectrum and the superpotential interactions are more readily
obtained in the perturbative heterotic string limit.
The merit of the nonperturbative limit will be in
trying to gain insight into phenomena which are intrinsically
nonperturbative in nature. Specifically, in trying to elucidate
the dynamical mechanism which is responsible for selecting a
specific string vacuum and the related topology changing transitions.
As they have brought to the fore the relevance of string compactifications
to the details of the Standard Model data, it is plausible that
the free fermionic models will also be instrumental to shed light
on these nontrivial issues. We will expand on this aspect in future
publications.

In this paper we make the first steps towards studying compactifications
of Ho\v rava--Witten theory on manifolds that are associated with the
realistic free fermionic models \cite{rffm}.
Heterotic M--theory compactifications
to four dimensions have been studied by Donagi {\it et al.}
\cite{daow,dlow}, on manifolds
that do not admit Wilson line breaking and yield $SU(5)$, $SO(10)$
or $E_6$ grand unified gauge groups \cite{daow},
as well as construction of
$SU(5)$ grand unified models that can be broken to the
Standard Model gauge group by Wilson line breaking \cite{dlow}.
In this paper
we extend the work of Donagi {\it et al.}, to the case of $SO(10)$ models
that allow Wilson line breaking. This entails the generalization of
the gauge bundle analysis of ref. \cite{dlow}
from $G=SU(2n+1)$ to $G=SU(2n)$ in the decomposition
of $E_8\supset G\times H$, where $H=SO(10)$ in our case.
As a result we advance the analysis
toward relating heterotic M--theory compactification to the
realistic free fermionic models. It should also be remarked,
as will be further discussed below and in future publications,
that the Calabi--Yau manifolds that are associated with the
free fermionic compactifications realize the structure
of the manifolds constructed by Donagi {\it et al.} More precisely,
they correspond to manifolds with fundamental group
${\bf Z}_2$, which is necessary for Wilson line breaking.
As our concrete example in this paper we discuss the breaking of the
$SO(10)$ symmetry by Wilson lines to $SU(5)\times U(1)$, which is
the flipped $SU(5)$ breaking pattern \cite{flipped}.

Our paper is organized as follows: in section \ref{revhetm} we review
the construction of the standard model of particle physics from M--theory.
Section \ref{hirsur} presents a classification of nonperturbative, heterotic
M--theory vacua of toroidally fibred Calabi--Yau 3--folds over
Hirzebruch surfaces $F_r$. Del Pezzo surfaces $dP_r$ have recently
been shown to exhibit a mysterious duality with toroidal compactifications
of M--theory \cite{vafa}; in section \ref{delpez} we extend our analysis to
the case of the del Pezzo surface $dP_3$. The overlap between these
constructions and the free fermionic models is highlighted in section
\ref{overlap}. Finally, section \ref{discx} contains
a discussion and conclusions.

\section{Review of standard models from heterotic M--theory}\label{revhetm}

This section summarizes the relevant information to construct the standard
model of elementary particles from heterotic M--theory. We follow ref.
\cite{dlow} closely.

\subsection{The anomaly--cancellation condition}\label{hwpict}

The 11--dimensional spacetime $M_{11}$ of M--theory is taken to be
\begin{equation}
M_{11}=M_4\times {S^1\over {\bf Z}_2}\times Z,
\label{spatime}
\end{equation}
where $M_4$ is 4--dimensional Minkowski spacetime, the compact eleventh
dimension
$S^1$ is modded out by the action of ${\bf Z}_2$, and $Z$ is a Calabi--Yau
(complex) 3--fold. There is a semistable holomorphic vector bundle $V_i$,
$i=1,2$ over the 3--fold $Z$ on the orbifold fixed plane at each of the two
fixed
points of the ${\bf Z}_2$--action on $S^1$. The structure group $G_i$ of
$V_i$ is a subgroup of $E_8$.

Fivebranes exist in the vacuum, which wrap holomorphic 2--cycles within
$Z$ and are parallel to the orbifold fixed planes. The fivebranes are
represented by a 4-form cohomology class $[W]$.

The Calabi--Yau 3--fold $Z$, the gauge bundles $V_i$ and the fivebranes are
subject to the cohomological constraint on $Z$
\begin{equation}
c_2(V_1)+c_2(V_2)+[W]=c_2(TZ),
\label{annox}
\end{equation}
where $c_2(V_i)$ is the second Chern class of the $i$--th gauge bundle and
$c_2(TZ)$ is the second Chern class of the holomorphic tangent bundle to $Z$.
Equation (\ref{annox}) above is referred to as the anomaly--cancellation
condition.

\subsection{Elliptically--fibred Calabi--Yau 3--folds}\label{efcytf}

An elliptically--fibred Calabi--Yau complex 3--fold $X$ consists of a base $B$,
which is a complex 2--fold, together with an analytic map
\begin{equation}
\pi:X\rightarrow B
\label{elix}
\end{equation}
such that the fibre $\pi^{-1}(b)$ at a generic point $b\in B$ is an
elliptic curve. We also require the existence of a section,
{\it i.e.}, an analytic map
\begin{equation}
\sigma : B\rightarrow X
\label{anysig}
\end{equation}
such that $\pi\circ \sigma={\bf 1}_B$.

Given this elliptic fibration, a line bundle ${\cal L}$ is defined over $B$
as the conormal bundle to the section in $X$.
The Calabi--Yau condition $c_1(TX)=0$ then implies that
\begin{equation}
c_1({\cal L})=c_1(B).
\label{calcond}
\end{equation}
The second Chern class $c_2(TX)$ has been found in ref. \cite{fmw} to be
\begin{equation}
c_2(TX)=c_2(B) + 11c_1(B)^2 + 12\sigma c_1(B).
\label{cduetx}
\end{equation}
The Calabi--Yau condition further requires that the base be a del Pezzo, an
Enriques, or a Hirzebruch surface, or a blowup of a Hirzebruch surface
\cite{mova}. For all these the Chern classes $c_1(B)$ and $c_2(B)$ are known.

\subsection{The spectral cover}\label{specov}

The method of spectral covers provides an effective construction of
holomorphic,
semistable gauge bundles $V_X$ on elliptically--fibred manifolds $X$
\cite{fmw, asian, bersha, dlow, bruzzo}. The gauge bundle breaks the observable
$E_8$ to $G\times H$.
For the gauge group $G=SU(n)$,
the required data are a divisor
$C\subset X$, called spectral cover, plus a line bundle ${\cal N}$ on $C$.
The divisor $C$ is an $n$--fold covering of the base $B$, {\it i.e.}, the
restriction $\pi_C:C\rightarrow B$ of the elliptic fibration $\pi$ is an
$n$--sheeted branched covering. We will not distinguish notationally
between $\pi$ and its restriction $\pi_C$. We will denote the
pullback and the
pushforward of (co)homology classes between $C$ and $B$ by $\pi^*$ and
$\pi_*$, respectively.

The requirement that
\begin{equation}
c_1(V_X)=0,
\label{cuno}
\end{equation}
imposed by $G=SU(n)$, implies that the line bundle ${\cal N}$ has a first
Chern class given by
\begin{equation}
c_1({\cal N})=-{1 \over 2} \left(c_1(C) - \pi^*c_1(B)\right) + \gamma.
\label{cenne}
\end{equation}
In the above equation,
\begin{equation}
c_1(C)=-n\sigma -\pi^*\eta
\label{cunoc}
\end{equation}
is the first Chern class of the surface $C$. The latter reads in homology
\begin{equation}
C=n\sigma + \pi^*\eta
\label{chom}
\end{equation}
and, since $C$ is an actual surface within $X$, we must impose the
condition that $\eta$ be an effective class in $B$. In cohomology,
$\eta$ is the first Chern class of a certain line bundle ${\cal M}$ on $B$,
\begin{equation}
\eta=c_1({\cal M}),
\label{uneta}
\end{equation}
and $\gamma\in H^2(C, {\bf Q})$ is a class whose pushforward to $H^2(B)$ vanishes,
\begin{equation}
\pi_*(\gamma)=0.
\label{pigam}
\end{equation}
The general solution to eqn. (\ref{pigam}) is
\begin{equation}
\gamma=\lambda\left(n\sigma - \pi^*\eta + n \pi^*c_1(B)\right)\cdot C,
\label{gax}
\end{equation}
where $\lambda$ is a rational parameter. Substituting eqns. (\ref{cunoc})
and (\ref{gax}) into (\ref{cenne}) we obtain
\begin{equation}
c_1({\cal N}) = n\left({1\over 2} + \lambda\right)\sigma +
\left({1\over 2} - \lambda\right)\pi^*\eta +
\left({1\over 2} + n\lambda\right)\pi^*\left(c_1(B)\right).
\label{clin}
\end{equation}
Now $c_1({\cal N})$ must be an integer class. This leads to various sets of
sufficient conditions on $\lambda$ and $\eta$ that ensure the integrality
of $c_1({\cal N})$. Following ref. \cite{dlow} we find that, when $n=2k$,
one such set is
\begin{equation}
\lambda\in{\bf Z}, \qquad \eta=c_1(B)\;{\rm mod}\; 2,
\label{req}
\end{equation}
where the modding is by an even element of $H^2(B, {\bf Z})$.
There are, however, alternative sets of sufficient conditions that ensure
the integrality of $c_1({\cal N})$. When $n=2k$, one such alternative set is
\begin{equation}
\lambda= {2m+1\over 2}, \, m\in {\bf Z}, \qquad c_1(B) \;
{\rm even}.
\label{altset}
\end{equation}
This alternative set will be analyzed extensively in section \ref{hirsur}.

\subsection{Torus--fibred Calabi--Yau 3--folds}\label{tftfold}

The breaking of $SO(10)$ to $SU(5)\times U(1)$ is done by means of
Wilson lines on the Calabi--Yau manifold $X$. Nontriviality of the Wilson
lines requires a nontrivial fundamental group $\pi_1(X)$. For the base
manifolds
$B$ enumerated above, only the Enriques surface gives rise to a nontrivial
$\pi_1(X)$. However, one proves that the Enriques surface is ruled out
for the reasons explained in ref. \cite{dlow}.

Over a base manifold given by a Hirzebruch surface $F_r$ or a del Pezzo
surface $dP_r$, a Calabi--Yau 3--fold $Z$ with $\pi_1(Z)={\bf Z}_2$ can be
constructed as the quotient of an elliptically--fibred Calabi--Yau $X$ by a
freely--acting involution $\tau_X$, {\it i.e.}, a map $\tau_X:X\rightarrow X$
satisfying $\tau^2_X={\bf 1}$. This construction necessitates a second section
$\zeta$,
\begin{equation}
\zeta : B\rightarrow X, \qquad \pi\circ\zeta={\bf 1}_B,
\label{necss}
\end{equation}
such that $\zeta + \zeta=\sigma$ under fibrewise addition. It turns out
that the involution preserves the fibration structure but exchanges the two
sections,
\begin{equation}
\tau_X(\zeta)=\sigma, \qquad \tau_X(\sigma)=\zeta,
\label{exchx}
\end{equation}
so the quotient space
\begin{equation}
Z=X/\tau_X
\label{zet}
\end{equation}
is torus fibred instead of elliptically fibred.
Let $\tau_B$ denote the involution of the base $B$ induced by the
involution $\tau_X$ of $X$. Since $\tau_X$ preserves the fibration $\pi$ of
$X$, it follows that $Z$ is a torus fibration over the base $B/\tau_B$.

Additional effects of the second section $\zeta$ are the following.

A curve of singularities appears in the section $\zeta$, which has to be blown
up
for $X$ to be smooth. The general elliptic fiber $F$ splits into two
spheres: the new fiber $N$, plus the proper transform of the singular
fiber, which is in the class $F-N$. The union of these new fibers $N$ over
the curve of singularities forms the exceptional divisor $E$. The latter
intersects
the spectral cover $C$ in such a way that, as a class in $X$,
\begin{equation}
E|_C=E\cdot C=4(\eta\cdot c_1(B)) N=\sum_iN_i
\label{exc}
\end{equation}
for some new curves $N_i\in H_2(C, {\bf Z})$. Their number is $4\eta\cdot
c_1(B)$,
and the class
\begin{equation}
{1\over 2}\sum_iN_i
\label{intx}
\end{equation}
has the important property of being integral.

The fivebranes physically wrap a holomorphic curve within the 3--fold.
This means that the cohomology class $[W]$ of the wrapped fivebranes must
be Poincar\'e--dual to the homology class of a set of (complex) curves in
the Calabi--Yau space, {\it i.e.}, $[W]$ must be effective as a
homology class. In general, the class of a curve in $H_2(X, {\bf Z})$ can
be written as
\begin{equation}
[W]=\sigma_*(\omega) + c(F-N) + dN,
\label{clascur}
\end{equation}
where $c,d$ are integers, $\omega$ is a class in the base $B$, and
$\sigma_*(\omega)$ is its pushforward to $X$ under the section
(\ref{anysig}).
One can show \cite{dlow} that a sufficient condition for $[W]$ to be effective
is
that $\omega$ be effective in $B$,  plus
\begin{equation}
c\geq 0, \qquad d\geq 0.
\label{efecond}
\end{equation}

{}Finally, the second Chern class of $X$ also gets modified by the presence of
the new
section $\zeta$,
\begin{equation}
c_2(TX)=12\sigma\pi^*c_1(B) + \left(c_2(B) + 11c_1^2(B)\right)(F-N)
+ \left(c_2(B) - c_1^2(B)\right)N.
\label{cduesec}
\end{equation}

\subsection{Bundles on torus--fibred 3--folds}\label{gbtf}

We will consider semistable, holomorphic vector bundles $V_Z$ over the 3--fold
$Z=X/\tau_X$. This corresponds to constructing supersymmetric vacua of the
gauge theory. Call $q$ the quotient map
\begin{equation}
q:X\rightarrow Z.
\label{qmap}
\end{equation}
Then Chern classes $c_i(TZ)$, $c_i(V_Z)$ of the tangent bundle and of the gauge
bundle
on $Z$ can be determined from those corresponding to $X$ by pushforward:
\begin{equation}
c_i(TZ)={1\over 2} q_*\left(c_i(TX)\right),\qquad
c_i(V_Z)={1\over 2} q_*\left(c_i(V_X)\right).
\label{pufo}
\end{equation}

The spectral--cover construction summarized in section \ref{specov} for
elliptically--fibred 3--folds $X$ may be adapted to torus--fibred 3--folds $Z$
\cite{dlow}. One continues to have a line bundle ${\cal N}$ on $C$ whose
first Chern class is given by eqns. (\ref{cenne}) -- (\ref{pigam}).
One can prove that the general solution to eqn. (\ref{pigam}) picks up
terms proportional to the new classes $N_i$, so the new $\gamma$ reads
\begin{equation}
\gamma=\lambda\left(n\sigma - \pi^*\eta + n \pi^*c_1(B)\right)\cdot C
+\sum_i\kappa_iN_i,
\label{nega}
\end{equation}
for arbitrary rational coefficients $\kappa_i$. As in section \ref{specov}
we have to impose the condition that $c_1({\cal N})$ (given by eqns.
(\ref{cenne}),
(\ref{nega}))
\begin{equation}
c_1({\cal N}) = n\left({1\over 2} + \lambda\right)\sigma +
\left({1\over 2} - \lambda\right)\pi^*\eta +
\left({1\over 2} + n\lambda\right)\pi^*\left(c_1(B)\right) +
\sum_i\kappa_iN_i
\label{clinx}
\end{equation}
be an integer class.  As in eqns. (\ref{req}) and (\ref{altset}), we find
various sets of sufficient conditions that ensure the integrality of
$c_1({\cal N})$. In the case when $n=2k$  one such set is
\begin{equation}
\lambda\in {\bf Z}, \qquad \eta=c_1(B)\;{\rm mod}\; 2,
\label{condlam}
\end{equation}
and the $\kappa_i$ can be either
all integer or all one--half an odd integer. Another set of sufficient conditions
when $n=2k$ is
\begin{equation}
\lambda= {2m+1\over 2}, \, m\in {\bf Z}, \qquad c_1(B) \;
{\rm even},
\label{altsetx}
\end{equation}
with the same requirements on the $\kappa_i$ as before.

{}Finally every semistable, holomorphic vector bundle $V_Z$ over the 3--fold
$Z=X/\tau_X$ can be pulled back to a bundle $V_X$ over
$X$ that is invariant under $\tau_X$, and conversely. This invariance
condition is expressed as
\begin{equation}
\tau_X^*(V_X)=V_X.
\label{invcon}
\end{equation}
An analysis of eqn. (\ref{invcon}) has been performed in ref.
\cite{dlow}. Here we will just quote the result that a set of necessary
conditions for $V_X$ to be invariant is
\begin{equation}
\tau_B(\eta)=\eta,\qquad \sum_i\kappa_i=\eta\cdot c_1(B).
\label{invcondx}
\end{equation}

\subsection{Observable and hidden sectors}\label{chcl}

In the Ho\v rava--Witten picture \cite{hw}, the gauge bundles on the
Calabi--Yau 3--folds appear as subbundles of the $E_8$ bundles on the
orbifold fixed planes. For simplicity, following \cite{dlow}, we will
place a trivial, unbroken  $E_8$--bundle on one of the two fixed planes
(the hidden sector). On the remaining orbifold plane (the observable sector)
the structure group will be an $SU(4)$ subgroup of $E_8$. Within $E_8$,
the group $H=SO(10)$ is the commutant of the structure group $G=SU(4)$.
Throughout this paper, when referring to an $SU(n)$--bundle on the 3--fold,
we refer to this particular $SU(4)$ whose commutant leaves behind an $SO(10)$
grand unified gauge theory in the observable sector.

Picking the trivial $E_8$--bundle on the hidden sector simplifies the
anomaly--cancellation condition (\ref{annox}) on the 3--fold $Z$ to
\begin{equation}
[W_Z]=c_2(TZ)-c_2(V_Z).
\label{simpz}
\end{equation}
Using eqn. (\ref{pufo}), this can be expressed as an anomaly--cancellation
condition on the covering 3--fold $X$,
\begin{equation}
[W_X]=c_2(TX)-c_2(V_X).
\label{simpx}
\end{equation}

Second Chern classes for $SU(n)$ gauge bundles over $X$ were computed in
refs. \cite{fmw, dlow, curio, salamanca}. For the second Chern class
in the case when $X$ admits only one section $\sigma$ we have
\begin{equation}
c_2(V_X)=\eta\sigma - {1\over 24}(n^3-n)c_1^2 -
{1\over 2}\left(\lambda^2-{1\over 4}\right)n\eta\left(\eta-nc_1\right),
\label{cvfmw}
\end{equation}
where $c_1$ denotes $c_1(B)$.
When $X$ admits two sections $\sigma$, $\zeta$ the result is \cite{dlow}
\begin{eqnarray}
&&c_2(V_X)=\sigma\cdot\pi^*\eta \cr
&-& \left[{1\over 24} (n^3-n)c_1^2
- {1\over 2}\left(\lambda^2-{1\over 4}\right) n\eta\left(\eta-nc_1\right) -
\sum_i\kappa_i^2\right] (F-N)\cr
&-& \left[{1\over 24} (n^3-n)c_1^2
- {1\over 2}\left(\lambda^2-{1\over 4}\right) n\eta\left(\eta-nc_1\right) -
\sum_i\kappa_i^2 + \sum_i \kappa_i\right] N,
\label{cvnoi}
\end{eqnarray}
which reduces to eqn. (\ref{cvfmw}) on setting $\kappa_i=0$.
Substituting eqn. (\ref{cvnoi}) into (\ref{simpx}) and denoting $c_2(B)$
simply by $c_2$, we can identify
the coefficients $c,d$ and the curve $\omega$ in eqn. (\ref{clascur}):
\begin{equation}
c=c_2 + \left({1\over 24} (n^3-n) + 11\right) c_1^2-
{1 \over 2}\left(\lambda^2-{1\over
4}\right)n\eta\left(\eta-nc_1\right)-\sum_i\kappa_i^2,
\label{cce}
\end{equation}
\begin{eqnarray}
d&=&c_2 + \left({1\over 24} (n^3-n) - 1\right) c_1^2 -
{1 \over 2}\left(\lambda^2-{1\over
4}\right)n\eta\left(\eta-nc_1\right)\cr
&-&\sum_i\kappa_i^2
+\sum_i\kappa_i,
\label{dde}
\end{eqnarray}
\begin{equation}
\omega=12c_1(B)-\eta.
\label{idenx}
\end{equation}

A physical requirement is the existence of three families of quarks and leptons
in the
observable sector. The net number $N_{\rm gen}$ of generations is related to
the third Chern
class of the gauge bundle through
\begin{equation}
N_{\rm gen}={1\over 2} \int_Z c_3(V_Z).
\label{nogen}
\end{equation}
Now the third Chern class reads \cite{curio, andreas}
\begin{equation}
c_3(V_Z)=2\lambda\sigma\eta\left(\eta - nc_1(B)\right).
\label{ctre}
\end{equation}
Using eqns. (\ref{pufo}), (\ref{nogen}), (\ref{ctre}) and integrating over the
fibre $F$ yields the requirement
\begin{equation}
\lambda\eta\left(\eta - n c_1(B)\right)=6.
\label{nogenx}
\end{equation}

In ref. \cite{bmr} a necessary condition has been worked out to ensure that
the commutant $H$ is actually the largest preserved subgroup of $E_8$,
for a given choice of $G$. In our case this condition reads
\begin{equation}
\eta\geq 4 c_1(B).
\label{berg}
\end{equation}

\subsection{Summary of rules}\label{suru}

The rules presented above allow one to construct
realistic particle physics vacua with $N=1$ supersymmetry, three families
of quarks and leptons and a grand unified gauge group $SO(10)$.
We refer to \cite{dlow} for the geometric conditions needed on the
elliptically--fibred 3--fold $X$ to produce a torus--fibred 3--fold $Z$.
Let us summarize the remaining rules before working out some explicit
examples.

a) Semistability condition: the spectral data $(C, {\cal N})$ specifying a
semistable, holomorphic vector bundle can be written, via eqns.
(\ref{cenne}), (\ref{chom}), (\ref{nega}) in terms of an effective class
$\eta\in H^2(B, {\bf Z})$ and coefficients $\lambda$, $\kappa_i$ satisfying
eqn. (\ref{condlam}),
\begin{equation}
\lambda\in {\bf Z}, \qquad \eta=c_1(B)\;{\rm mod}\; 2,
\label{condlamx}
\end{equation}
(and the $\kappa_i$ either all integer or all half an odd integer)
or eqn. (\ref{altsetx}),
\begin{equation}
\lambda= {2m+1\over 2}, \, m\in {\bf Z}, \qquad c_1(B) \;
{\rm even},
\label{altsetxz}
\end{equation}
(with the same requirements on the $\kappa_i$ as above).

b) Involution conditions: for a vector bundle $V_X$ on $X$ to
descend to a vector bundle $V_Z$ on $Z$ it is necessary that
\begin{equation}
\tau_B(\eta)=\eta,\qquad \sum_i\kappa_i=\eta\cdot c_1(B).
\label{invcondxz}
\end{equation}

c) Effectiveness condition: a sufficient condition for $[W]$ in eqn.
(\ref{clascur}) to be an effective class is
\begin{equation}
12c_1(B)\geq \eta,\qquad c\geq 0,\qquad d\geq 0.
\label{efectdonc}
\end{equation}

d) Commutant condition: for the gauge group $SO(10)$ this condition reads
\begin{equation}
\eta\geq 4 c_1(B).
\label{bergx}
\end{equation}

e) Three--family condition:
\begin{equation}
\lambda\eta\left(\eta - n c_1(B)\right)=6.
\label{nogenxz}
\end{equation}

\section{Vacua over Hirzebruch surfaces $F_r$}\label{hirsur}

We take the base manifold $B$ to be the Hirzebruch surface $F_r$,
$r\geq 0$ \cite{calgsurf}.
The latter is a ${\bf CP}^1$--fibration over ${\bf CP}^1$.
A basis for $H_2(F_r, {\bf Z})$ composed of effective classes
is given by the class of the base ${\bf CP}^1$,
denoted $S$, plus the class of the fibre ${\bf CP}^1$,
denoted $E$. Their intersections are
\begin{equation}
S\cdot S= -r, \qquad S\cdot E=1,\qquad
E\cdot E=0.
\label{fintmat}
\end{equation}
All effective curves in $F_r$ are linear combinations of $S$ and
$E$ with nonnegative coefficients. The Chern classes of $F_r$ are
\begin{equation}
c_1(F_r)=2S + (r+2) E, \qquad c_2(F_r)=4.
\label{fcla}
\end{equation}
It is proved in ref. \cite{dlow} that, over the base $F_r$, one can construct
torus--fibred Calabi--Yau 3--folds $Z$ whose fundamental group is ${\bf Z}_2$
when $r=0,2$. For those allowed values of $r$, any class $\eta\in H_2(F_r, 
{\bf Z})$ is $\tau_B$--invariant \cite{dlow}.
In what follows we will work with an arbitrary allowed value of $r$.
Let us write $\eta\in H_2(F_r, {\bf Z})$ as
\begin{equation}
\eta=sS + e E
\label{lewr}
\end{equation}
for some integers $s, e$ to be determined imposing the conditions summarized
in section \ref{suru}.

{}For the semistability condition we have a choice. Either we impose eqn.
(\ref{condlamx}), which  implies that
\begin{equation}
\lambda\in {\bf Z},\qquad s\; {\rm even},\qquad e-r\; {\rm even},
\label{parz}
\end{equation}
or we impose condition (\ref{altsetxz}),
\begin{equation}
\lambda={2m+1\over 2},\; m\in {\bf Z},\qquad r\; {\rm even}.
\label{pparc}
\end{equation}
The involution conditions (\ref{invcondxz}) are
\begin{equation}
\sum_i\kappa_i=2s+2e-sr,
\label{invcondxzz}
\end{equation}
while the effectiveness conditions (\ref{efectdonc}) read
\begin{equation}
24\geq s,\qquad 12r+24 \geq e,
\label{togw}
\end{equation}
\begin{equation}
\sum_i \kappa_i^2 \leq 112 + {3\over \lambda} - 12\lambda
\label{cczz}
\end{equation}
and
\begin{equation}
\sum_i \kappa_i^2\leq 16 + \sum_i \kappa_i + {3\over \lambda} - 12\lambda
\label{ddzz}.
\end{equation}
The commutant condition (\ref{bergx}) requires
\begin{equation}
s\geq 8, \qquad e\geq 4r+8.
\label{bergz}
\end{equation}
{}Finally we analyze the three--family condition (\ref{nogenxz}):
\begin{equation}
-rs^2+4rs+2es-8e-8s = 6/\lambda.
\label{sugw}
\end{equation}

Now the left--hand side of eqn. (\ref{sugw}) is always an even integer,
whatever our choice for the semistability condition (eqn. (\ref{parz}) or
(\ref{pparc})).
If we make the choice (\ref{parz}), $\lambda$ can only take the values
$\pm 1,\pm 2,\pm 3,\pm 6$. Parity rules out the values $|\lambda|=2,6$ and
allows
$|\lambda| = 1, 3$. We denote this family of solutions as class A:
\begin{equation}
{\rm class}\; {\rm A:}\qquad s\; {\rm even},\qquad e-r\; {\rm even},\qquad
\lambda = \pm 1,\pm 3.
\label{clasa}
\end{equation}
Choosing (\ref{pparc}) instead, then integrality of $6/\lambda= 12/(2m+1)$
restricts $m$
to $-2,-1,0,1$, {\it i.e.}, $|\lambda|=1/2, 3/2$. This we call class B:
\begin{equation}
{\rm class}\; {\rm B:}\qquad r\; {\rm even},\qquad
\lambda = \pm 1/2,\pm 3/2.
\label{clasb}
\end{equation}

Next we solve the three--family condition (\ref{sugw}) for $e$,
assuming a given value for $s$. This gives $e$ as a function of $r$ and
$\lambda$:
\begin{equation}
e(r;\lambda)={1\over 2s-8}\left(rs^2-4rs+8s +{6\over \lambda}\right).
\label{reso}
\end{equation}
{}For each value of $8\leq s\leq 24$ and the corresponding appropriate choice
for $\lambda$
we present below the solutions for $e(r;\lambda)$. We indicate it whenever the
solution
is not an integer for any allowed value of $\lambda$.

Class A)
\begin{eqnarray}
s&=&8,\qquad\,\;e(r;\lambda) = 4r + 8 + 3/4\lambda\notin {\bf Z}\cr
s&=&10,\qquad e(r;\lambda) = 5r + 20/3 + 1/2\lambda\notin {\bf Z}\cr
s&=&12,\qquad e(r;\lambda) = 6r + 6 + 3/8\lambda\notin {\bf Z}\cr
s&=&14,\qquad e(r;\lambda) = 7r + 28/5 + 3/10\lambda\notin {\bf Z}\cr
s&=&16,\qquad e(r;\lambda) = 8r + 16/3 + 1/4\lambda\notin {\bf Z}\cr
s&=&18,\qquad e(r;\lambda) = 9r + 36/7 + 3/14\lambda\notin {\bf Z}\cr
s&=&20,\qquad e(r;\lambda) = 10r + 5 + 3/16\lambda \notin {\bf Z}\cr
s&=&22,\qquad e(r;\lambda) = 11r + 44/9 + 1/6\lambda \notin {\bf Z}\cr
s&=&24,\qquad e(r;\lambda) = 12r + 24/5 + 3/20\lambda \notin {\bf Z}.
\end{eqnarray}
There are no integer solutions in this class. We conclude that class A
contains no vacua over the Hirzebruch surface $F_r$,
for any allowed value of $r$.

Class B)
\begin{eqnarray}
s&=&8,\qquad\,\; e(r;\lambda) = 4r + 8 + 3/4\lambda\notin {\bf Z}\cr
s&=&9,\qquad\,\; e(r;\lambda=-1/2) = 9r/2 + 6\cr
s&=&10,\qquad e(r;\lambda=3/2) = 5r + 7\cr
s&=&11,\qquad e(r;\lambda=-3/2) = 11r/2 +6\cr
s&=&12,\qquad e(r;\lambda) = 6r + 6 + 3/8\lambda\notin {\bf Z}\cr
s&=&13,\qquad e(r;\lambda=3/2) = 13r/2 + 6\cr
s&=&14,\qquad e(r;\lambda=-1/2) = 7r + 5\cr
s&=&15,\qquad e(r;\lambda=1/2) = 15r/2 + 6\cr
s&=&16,\qquad e(r;\lambda) = 8r + 16/3 + 1/4\lambda\notin {\bf Z}\cr
s&=&17,\qquad e(r;\lambda) = 17r/2 + 68/13 + 3/13\lambda\notin{\bf Z}\cr
s&=&18,\qquad e(r;\lambda=-3/2) = 9r + 5\cr
s&=&19,\qquad e(r;\lambda) = 19r/2 + 76/15 + 1/5\lambda \notin {\bf Z}\cr
s&=&20,\qquad e(r;\lambda) = 10r + 5 + 3/16\lambda \notin {\bf Z}\cr
s&=&21,\qquad e(r;\lambda) = 21r/2 + 84/17 + 3/17\lambda \notin {\bf Z}\cr
s&=&22,\qquad e(r;\lambda=3/2) = 11r + 5\cr
s&=&23,\qquad e(r;\lambda) = 23r/2 + 92/19 + 3/19\lambda \notin {\bf Z}\cr
s&=&24,\qquad e(r;\lambda) = 12r + 24/5 + 3/20\lambda \notin {\bf Z}.
\label{tabula}
\end{eqnarray}
This class does lead to integer solutions to the three--family equation
(\ref{reso}).
In all cases the commutant condition (\ref{bergz}) is satisfied,
as well as eqn. (\ref{togw}) about the effectiveness of $\omega$ in $B$.

We can now return to eqn. (\ref{clascur}) and write explicit expressions for
the homology class $[W]$ that is being wrapped by the fivebranes on the
torus--fibred Calabi--Yau 3--fold $Z$.  We have for the class $\omega$
\begin{equation}
\omega=(24-s)S + (12r+24-e)E,
\label{havom}
\end{equation}
and for the coefficients $c$, $d$,
\begin{equation}
c=112+{3\over \lambda} -12\lambda -\sum_i\kappa_i^2,
\label{ece}
\end{equation}
\begin{equation}
d=16+{3\over \lambda} - 12\lambda +\sum_i\kappa_i -\sum_i\kappa_i^2.
\label{ede}
\end{equation}
Evaluating eqns. (\ref{havom}), (\ref{ece}) and (\ref{ede}) at the integer
solutions for $e(r;\lambda)$ tabulated in eqn. (\ref{tabula}), we arrive
at the following vacua $[W]$. Every allowed choice of $r$, plus every
choice of the rational coefficients $\kappa_i$ subject to the conditions
indicated
in each case, gives rise to a different class $[W]$:

$\bullet$ $s=9$: $\sum_i\kappa_i=30$ and $\sum_i\kappa_i^2\leq 46$,
\begin{equation}
[W]=\sigma_*\left(15S+\left({15\over 2}r + 18\right) E\right)+
(112-\sum_i\kappa_i^2)(F-N) +
(46-\sum_i\kappa_i^2)N.
\label{snove}
\end{equation}

$\bullet$ $s=10$: $\sum_i\kappa_i=34$ and $\sum_i\kappa_i^2\leq 34$,
\begin{equation}
[W]=\sigma_*\left(14 S + (7r + 17) E\right) +
(96-\sum_i\kappa_i^2) (F-N) +
(34 -\sum_i\kappa_i^2)N.
\label{sdiec}
\end{equation}

$\bullet$ $s=11$: $\sum_i\kappa_i=34$ and $\sum_i\kappa_i^2\leq 66$,
\begin{equation}
[W]=\sigma_*\left(13S+\left({13\over 2} r + 18\right)E\right) +
(128 - \sum_i \kappa_i^2)(F-N) +
(66 - \sum_i\kappa_i^2)N.
\label{sundici}
\end{equation}

$\bullet$ $s=13$:  $\sum_i\kappa_i=38$ and $\sum_i\kappa_i^2\leq 38$,
\begin{equation}
[W]=\sigma_*\left(11 S + \left({11\over 2} r + 18 \right) E\right) +
(96 - \sum_i\kappa_i^2)(F-N) +
(38 - \sum_i\kappa_i^2)N.
\label{sdodici}
\end{equation}

$\bullet$ $s=14$: $\sum_i\kappa_i=38$ and $\sum_i\kappa_i^2\leq 54$,
\begin{equation}
[W]=\sigma_*\left(10S + (5r + 19)E\right) +
(112 - \sum_i\kappa_i^2)(F-N) +
(54 -\sum_i\kappa_i^2)N.
\label{squatto}
\end{equation}

$\bullet$ $s=15$: $\sum_i\kappa_i=42$ and $\sum_i\kappa_i^2\leq 58$,
\begin{equation}
[W]=\sigma_*\left(9S + \left({9\over 2}r+18\right)E\right) +
(112-\sum_i\kappa_i^2)(F-N) +
(58-\sum_i\kappa_i^2)N.
\label{squindici}
\end{equation}

$\bullet$ $s=18$: $\sum_i\kappa_i=46$ and $\sum_i\kappa_i^2\leq 78$,
\begin{equation}
[W]=\sigma_*\left(6 S + (3r+19)E\right) +
(128-\sum_i\kappa_i^2)(F-N) +
(78-\sum_i\kappa_i^2)N.
\label{sdiciotto}
\end{equation}

$\bullet$ $s=22$: $\sum_i\kappa_i=54$ and $\sum_i\kappa_i^2\leq 54$,
\begin{equation}
[W]=\sigma_*\left(2S + (r+19)E\right) +
(96-\sum_i\kappa_i^2) (F-N) +
(54-\sum_i\kappa_i^2)N.
\label{sventidue}
\end{equation}

Each one of the above classes represents a nonperturbative vacuum
of an $SO(10)$ grand unified theory of particle physics.

\section{Vacua over the del Pezzo surface $dP_3$}\label{delpez}

As our next example we choose the base $B$ to be the del Pezzo surface
$dP_3$ (for nice reviews on del Pezzo surfaces see \cite{vafa, dlow}).
The latter can be thought of as complex projective space ${\bf CP}^2$,
blown up at three points. A basis of $H_2(dP_3, {\bf Z})$ composed entirely
of effective classes is given by the hyperplane class $l$,
plus three exceptional divisors $E_i$, $i=1,2,3$. Their intersections are
\begin{equation}
l\cdot l=1, \qquad E_i\cdot E_j=-\delta_{ij}, \qquad E_i\cdot l=0.
\label{dpint}
\end{equation}
The first and second Chern classes are given by
\begin{equation}
c_1(dP_3)=3l-E_1-E_2-E_3, \qquad c_2(dP_3)=6.
\label{ccdpt}
\end{equation}
It is proved in ref. \cite{dlow} that, over the base $dP_3$,
one can construct torus--fibred Calabi--Yau 3--folds $Z$ whose fundamental
group $\pi_1(Z)$ is ${\bf Z}_2$. It is convenient to consider the
independent curves
\begin{eqnarray}
{M}_1&=&l+E_1-E_2-E_3\cr
{M}_2&=&l-E_1+E_2-E_3\cr
{M}_3&=&l-E_1-E_2+E_3,
\label{acurv}
\end{eqnarray}
whose intersection matrix is
\begin{equation}
{M}_1\cdot {M}_1={M}_2\cdot {M}_2={M}_3\cdot
{M}_3= -2,\qquad
{M}_1\cdot {M}_2={M}_1\cdot {M}_3={M}_2\cdot
{M}_3= 2.
\label{inta}
\end{equation}
It turns out that the ${M}_j$ are effective classes and that they satisfy
$\tau_B({M}_j)=M_j$. They also generate all other
$\tau_B$--invariant curves, so the most general $\tau_B$--invariant
class $\eta\in H_2(dP_3, {\bf Z})$ is a linear combination
\begin{equation}
\eta=m_1 M_1 + m_2 M_2 + m_3 M_3,
\label{lincomv}
\end{equation}
for some arbitrary integer coefficients $m_j$ to be determined imposing the
requirements summarized in section \ref{suru}. The first Chern class
$c_1(dP_3)$ reads, in terms of the $M_j$,
\begin{equation}
c_1(dP_3) = M_1 + M_2 + M_3.
\label{cunotre}
\end{equation}

Next we impose the rules summarized in section \ref{suru}. For the
semistability condition we cannot impose eqn. (\ref{altsetxz}) in view of
$c_1(dP_3)$, so we have (\ref{condlamx}) instead:
\begin{equation}
\lambda\in {\bf Z},\qquad m_j\; {\rm odd}, j=1,2,3,
\label{sem}
\end{equation}
and either all the $\kappa_i$ integer or all half an odd integer.
With our choice (\ref{lincomv}) for $\eta$, the involution condition
(\ref{invcondxz}) reduces to
\begin{equation}
\sum_i\kappa_i=2(m_1+m_2+m_3).
\label{olina}
\end{equation}
The effectiveness conditions (\ref{efectdonc}) read
\begin{equation}
m_j\leq 12,\qquad j=1,2,3
\label{mdoc}
\end{equation}
and
\begin{equation}
\sum_i\kappa_i^2\leq 87 + {3\over \lambda} - 12\lambda,
\label{cdptrex}
\end{equation}
\begin{equation}
\sum_i\kappa_i^2\leq 15 + {3\over \lambda} - 12\lambda + \sum_i\kappa_i.
\label{ddptrex}
\end{equation}
The commutant condition (\ref{bergx}) requires
\begin{equation}
m_j\geq 4, \qquad j=1,2,3.
\label{berglu}
\end{equation}
{}Finally, the three--family condition (\ref{nogenxz}) is expressed as
\begin{equation}
-m_1^2 - m_2^2 - m_3^2 + 2(m_1m_2 + m_1m_3 + m_2m_3)
-4(m_1 + m_2 + m_3) = 3/\lambda.
\label{wurst}
\end{equation}
Given that the $m_j$ are odd, the left--hand side of the above is always odd,
so the allowed values for $\lambda$ are $\pm 1, \pm 3$.

Odd integer solutions to eqn. (\ref{wurst}) in the range $4\leq m_j\leq 12$,
for (at least) one of the allowed values for $\lambda$, correspond to
nonperturbative vacua of heterotic M--theory, compactified on a
torus--fibred Calabi--Yau 3--fold over a $dP_3$ surface, with $SO(10)$
as GUT group and 3 families of chiral matter. One can see that
for $\lambda=\pm 1, \pm 3$ there are no solutions in the required range,
and hence no vacua.
A simple proof of this fact is as follows. Assume that all solutions $m_j$
are equal to a  certain value $m$: $m_1=m_2=m_3=m$, then solve
eqn. (\ref{wurst}) for $m$. This gives $m=2\pm \sqrt{4+ 1/\lambda}$, which
for $\lambda = \pm 1, \pm 3$ is not an integer. Next assume that two
solutions are equal, say $m_1=m_2=m$, and that $m_3\neq m_1$. Then eqn.
(\ref{wurst}) becomes linear in $m$, but its solutions when $m_3=5,7,9,11$
are never an odd integer for $\lambda =\pm 1,\pm 3$. Finally, when
all the $m_j$ are pairwise different, one can easily check numerically
for $m_j=5,7,9,11$ that there is no solution for $\lambda = \pm 1, \pm 3$.

\section{Overlap with the free fermionic models}\label{overlap}

In this section we elaborate briefly on the overlap with the free fermionic
models. Amazingly enough, the structure of the manifolds constructed
by Donagi \etal, up to the imposition of the three generation
condition, precisely coincides with the structure of the
manifold that underlies the free fermionic models.

In the free fermionic formalism \cite{fff} a model is specified in terms
of a set of boundary condition basis vectors and one--loop
GSO projection coefficients. These fully determine the
partition function of the models, the spectrum and
the vacuum structure. The three generation models
of interest here are constructed in two stages. The first
corresponds to the NAHE set of boundary basis vectors
$\{{\bf1},S,b_1,b_2,b_3\}$ \cite{nahe}. The second consists
of adding to the NAHE set three additional boundary
condition basis vectors, typically denoted $\{\alpha,\beta,\gamma\}$.
The gauge group at the level of the NAHE set is $SO(6)^3\times
SO(10)\times E_8$, which is broken to $SO(4)^3\times U(1)^3\times
SO(10)\times SO(16)$ by the vector $2\gamma$. Alternatively,
we can start with an extended NAHE set $\{{\bf1},S,\xi_1,\xi_2,b_1,
b_2\}$, with $\xi_1={\bf1}+b_1+b_2+b_3$. The set $\{{\bf1},S,\xi_1,
\xi_2\}$ produces a toroidal Narain model with $SO(12)\times
E_8\times E_8$ or $SO(12)\times SO(16)\times SO(16)$ gauge
group depending on the GSO phase $c({\xi_1\atop\xi_2})$.
The basis vectors $b_1$ and $b_2$ then break $SO(12)\rightarrow
SO(4)^3$, and either $E_8\times E_8\rightarrow E_6\times U(1)^2\times E_8$
or $SO(16)\times SO(16)\rightarrow SO(10)\times U(1)^3\times SO(16)$.
The vectors $b_1$ and $b_2$ correspond to ${\bf Z}_2\times {\bf Z}_2$ orbifold
modding. The three sectors $b_1$, $b_2$ and $b_3$ correspond to
the three twisted sector of the ${\bf Z}_2\times {\bf Z}_2$ orbifold,
with each producing eight generations in the ${\bf27}$, or
${\bf16}$, representations
of $E_6$, or $SO(10)$, respectively. In the case of $E_6$ the untwisted
sector produces an additional $3\times ({\bf 27}+{\overline{\bf 27}})$,
whereas in the $SO(10)$ model
it produces $3\times({\bf 10}+{\overline{\bf 10}})$.
Therefore, the Calabi--Yau manifold that corresponds to the
${\bf Z}_2\times {\bf Z}_2$ orbifold at the free fermionic point
in the Narain moduli space has $(h_{11},h_{21})=(27,3)$.

To note the overlap with the construction of Donagi \etal
we construct the ${\bf Z}_2\times {\bf Z}_2$ at a generic point in the
moduli space. For this purpose, let us first start with the compactified
torus $T^2_1\times T^2_2\times T^2_3$  parameterized by
three complex coordinates $z_1$, $z_2$ and $z_3$,
with the identification
\begin{equation}
z_i=z_i + 1~~~~~~~~~~;~~~~~~~~~~z_i=z_i+\tau_i,
\label{t2cube}
\end{equation}
where $\tau$ is the complex parameter of each torus
$T^2$.
We consider ${\bf Z}_2$ twists and possible shifts of order
two:
\begin{equation}
z_i~\rightarrow~(-1)^{\epsilon_i}z_i+{1\over 2}\delta_i,
\label{z2twistanddance}
\end{equation}
subject to the condition that $\Pi_i(-1)^{\epsilon_i}=1$.
This condition insures that the holomorphic three--form
$\omega=dz_1\wedge dz_2\wedge dz_3$ is invariant under the ${\bf Z}_2$ twist.
Under the identification $z_i\rightarrow-z_i$, a single torus
has four fixed points at
\begin{equation}
z_i=\{0,1/2,\tau/2,(1+\tau)/2\}.
\label{fixedtau}
\end{equation}
The first model that we consider is produced
by the two ${\bf Z}_2$ twists
\begin{eqnarray}
&& \alpha:(z_1,z_2,z_3)\rightarrow(-z_1,-z_2,~~z_3)\cr
&&  \beta:(z_1,z_2,z_3)\rightarrow(~~z_1,-z_2,-z_3).
\label{alphabeta}
\end{eqnarray}
There are three twisted sectors in this model, $\alpha$,
$\beta$ and $\alpha\beta=\alpha\cdot\beta$, each producing
16 fixed tori, for a total of 48. The untwisted sector
adds three additional K\"ahler and complex deformation
parameters producing in total a manifold with $(h_{11},h_{21})=(51,3)$.
We refer to this model as $X_1$.
This manifold admits an elliptic fibration over a base $F_0={\bf CP}^1\times
{\bf CP}^1$. This can be seen from the  Borcea--Voisin classification
of elliptically fibered Calabi--Yau manifolds \cite{borceaviosin}
and from ref. \cite{mukhi}.We emphasize that in the discussion here
our intention is not to reproduce the generic M--theory elliptic
fibration model, but rather to indentify the overlap with the
specific geometrical structures that underly the realistic
free fermionic models. We further remark that here we only discuss
the correspondence with the compactified geometry and it would
be of further interest to elucidate how the gauge
bundles are related in the fermionic and the M-theory models.

Next we consider the model generated by the ${\bf Z}_2\times {\bf Z}_2$
twists in (\ref{alphabeta}), with the additional shift
\begin{equation}
\gamma:(z_1,z_2,z_3)\rightarrow(z_1+{1\over2},z_2+{1\over2},z_3+{1\over2}).
\label{gammashift}
\end{equation}
This model again has fixed tori from the three
twisted sectors $\alpha$, $\beta$ and $\alpha\beta$.
The product of the $\gamma$ shift in (\ref{gammashift})
with any of the twisted sectors does not produce any additional
fixed tori. Therefore, this shift acts freely.
Under the action of the $\gamma$ shift, half
the fixed tori from each twisted sector are paired.
Therefore, the action of this shift is to reduce
the total number of fixed tori from the twisted sectors
by a factor of $1/2$. Consequently, in this model
$(h_{11},h_{21})=(27,3)$. This model therefore
reproduces the data of the ${\bf Z}_2\times {\bf Z}_2$ orbifold
at the free-fermion point in the Narain moduli space.
We refer to this model as $X_2$.

The manifold  $X_1$ therefore coincides with the  manifold $X$
of Donagi \etal, the  manifold  $X_2$ coincides with the
 manifold $Z$, and the $\gamma$--shift in eq. (\ref{gammashift})
coincides with the freely acting involution $\tau_X$ in eqn.
(\ref{exchx},\ref{zet}). Thus, the free
fermionic models admit precisely the structure of the
Calabi-Yau manifolds considered in ref. \cite{dlow}.

\section{Discussion and conclusions}\label{discx}

The role of the freely acting shift discussed in the previous
section and employed in sections \ref{hirsur}, \ref{delpez}
is to produce a manifold which is not simply connected,
with $\pi_1(Z)={\bf Z}_2$.
This enables the use of Wilson lines
to break the $SO(10)$ symmetry to one of its subgroups.
For example, up to $SO(10)$ automorphisms, the only generator of
$SO(10)$ that leaves $SU(5)$ unbroken is given by \cite{campbell}
\begin{equation}
-iH=\left(\matrix{0&-1& & & & & & \cr
                  1&~0& & & & & & \cr
                   &  &.& & & & & \cr
                   &  & &.& & & & \cr
                   &  & & &.& & & \cr
                   &  & & & &0&-1 \cr
                   &  & & & &1&~0 \cr}\right).
\label{wilsonmatrix}
\end{equation}
Other breaking patterns, that break the
$SO(10)$ symmetry to one of its other subgroups, as in ref. \cite{rffm},
are possible, and in some cases by utilizing two independent
Wilson lines \cite{rffm}. Here the third Chern class counts
the number of ${\bf 16}$ minus ${\overline {\bf 16}}$ representations of
$SO(10)$. The net number $N_{\rm gen}=3$ then contains the three
chiral Standard Model generations. The additional ${\bf 10}\oplus{\overline
{\bf 10}}$ representations of $SU(5)$ needed to break the $SU(5)\times U(1)$
symmetry arise from additional ${\bf 16}\oplus{\overline {\bf 16}}$
representations
that are obscured from $N_{\rm gen}$ but, in general, appear
in the physical spectrum. The adjoint ${\bf 248}$ representation of
$E_8$ decomposes as ${\bf 248}=({\bf 45},{\bf 1})+({\bf 1},{\bf 15})+
({\bf 10},{\bf 6})+({\bf 16},{\bf 4})+({\overline {\bf 16}},{\overline {\bf
4}})$
under $SO(10)\times SU(4)$. As is the case in the example of the
free fermionic string models the $({\bf 10},{\bf 6})$ component in this
decomposition can produce the $SO(10)$ vectorial representation,
which contains the electroweak Higgs multiplets of the Minimal
Supersymmetric Standard Model, and may produce the fermion masses
from the ${\bf 16\cdot16\cdot10}$ superpotential term.

The question, however, is the utility of the sophisticated
mathematical tools employed in this paper. It is rather
plausible that details of the
massless spectrum and fermion masses
can be more readily obtained by using the conformal field
theory based formulations of the perturbative string limits.
Indeed, this lesson we can already infer from
heterotic string studies that yielded detailed
fermion mass textures, addressing issues like Cabibbo
mixing and fermion mass hierarchy \cite{fm}. It is
doubtful that the geometry based formalism
can explore a similar level of detail.

As we saw in sections \ref{hirsur} and \ref{delpez}, one
utility of the geometrical approach is in classification
of the available geometries by the phenomenological
criteria that they allow or disallow. While
Hirzebruch surfaces $F_r$, $r=0,2$, provide three generation solutions
with $SO(10)$ symmetry that may be broken by Wilson lines,
the del Pezzo surface $dP_3$ does not admit such solutions,
within the classes that we have analyzed here. Whether this observation
can be generalized to any del Pezzo surface $dP_r$ is an
interesting question that will be addressed in a forthcoming
publication \cite{inprep}. Thus, the geometrical insight provides a
tool to classify the manifolds according to very
basic phenomenological criteria.

More importantly, however, it is apparent that the
power of the complex manifold analysis, in the
particle phenomenology context, will be revealed
in trying to elucidate basic issues like the
topology changing transitions and vacuum selection.
In this respect we can promote the following view
of the utility of the M--theory picture that emerged
in recent years. It is now conjectured that the
different string theories are limits of one single, still
elusive, more basic theory. Each limit can then be utilized
to probe the properties of the more fundamental theory,
or its properties that may pertain to the observed particle
physics phenomenology. Thus, one limit may be utilized to
extract classes of manifolds that possess appealing
phenomenological characteristics, whereas another
limit may be useful to investigate dynamical transitions
between nearby manifolds.

In this context,
it has long been argued that the ${\bf Z}_2\times {\bf Z}_2$ orbifold
compactification naturally gives rise to three chiral generations
\cite{foc}. Modding
the 6--dimensional compactified space by
the ${\bf Z}_2\times {\bf Z}_2$ orbifold
projection produces three twisted sectors. In free fermionic string models
that are connected to the ${\bf Z}_2\times {\bf Z}_2$ orbifold each twisted
sector gives rise to one chiral generation. Thus, in these models the
existence of three generations is correlated with the underlying
geometrical structure.

The existence of three twisted sectors is a generic property  of the
${\bf Z}_2\times {\bf Z}_2$ orbifold of a 6--dimensional compactified space.
At a generic point of the compactified space one can take the moduli space
to be that of $T^2\times T^2\times T^2$ yielding a model with $(h_{21},
h_{11})=(51,3)$. At the free fermionic point the symmetry is enhanced,
producing an $SO(12)$ lattice. Taking the ${\bf Z}_2\times {\bf Z}_2$ orbifold
of the $SO(12)$ lattice then produces a model with $(h_{21},
h_{11})=(27,3)$. Each of these models has three twisted sectors with
16 and 8 fixed points in their respective twisted sectors.
Depending on a discrete torsion phase which commutes with
the ${\bf Z}_2\times {\bf Z}_2$ orbifold projection we can set the
4--dimensional nonabelian observable symmetry to be $SO(10)$ or $E_6$.
As we discussed in this paper, in order to allow Wilson line breaking
of the nonabelian $SO(10)$ or $E_6$ GUT symmetry, a Calabi--Yau manifold
has to be nonsimply connected. From refs. \cite{dlow, curio, efn3}
we learned that a simple way to achieve this is by modding a simply
connected Calabi--Yau manifold by a freely--acting involution, the latter
being ${\bf Z}_2$ in the models studied here and in \cite{dlow, curio, efn3}.
Now, the $(51,3)$ and $(27,3)$ ${\bf Z}_2\times {\bf Z}_2$ orbifold
compactifications are connected by precisely such a freely--acting involution.
Thus, what is remarkable is that precisely at the free fermionic point in
the moduli space, we find that the model naturally accommodates three
generations due to the ${\bf Z}_2\times {\bf Z}_2$ orbifold structure,
with the desirable $SO(10)$ Grand Unified symmetry, while at the same time
it allows for the inclusion of the Wilson line to break the GUT symmetry,
due to the freely--acting involution.

Thus, we see that precisely at the free fermionic point in the string
moduli space, some of the needed ingredients coalesce to produce the
phenomenologically required features. This remarkable coincidence
is, however, valid at weak coupling. The issue to understand is whether it
remains valid in the strong coupling regime. The compactification of Ho\v
rava--Witten theory, as discussed here, provides the means to investigate
such questions.

\bigskip
\leftline{\large\bf Acknowledgments}
\medskip

A.E.F. thanks Adam Ritz for discussions during the initial
stages of this project.
J.M.I. would like to thank U. Bruzzo and SISSA, Trieste (Italy) for
discussions and hospitality during the early stages of this work.
Support from PPARC (grants PPA/A/S/1998/00179 and
PPA/G/O/2000/00469) and DGICYT (grant PB 96-0756)
is acknowledged.


\begin{thebibliography}{99}

\bibitem{mreviews} For reviews and references, see {\it e.g.},\\
                        M.J. Duff, hep-th/9611203;\\
                        P.K. Townsend, hep-th/9612121;\\
                        A. Sen, hep-th/9802051;\\
                        B.A. Ovrut, hep-th/0201032.


\bibitem{nexperiments} SuperKamiokande collaboration,
                   Y. Fukuda \etal, \PLB{436}{1998}{33};\\
               SNO collaboration, Q.R. Ahmad \etal, \PRL{87}{2001}{071301}.

\bibitem{gcu} K.R. Dienes and \AEF, \NPB{457}{1995}{409};\\
              K.R. Dienes, \AEF~and J. March--Russell, \NPB{467}{1996}{44};\\
              D.M. Ghilencea and G.G. Ross, \NPB{606}{2001}{101}.
\bibitem{nonso10} A. Font, L.E. Ibanez, F. Quevedo and A. Sierra,
                                \NPB{331}{1990}{421};\\
            J.A. Casas and C. Mu{\~n}oz, \PLB{214}{1988}{63};\\
            S. Chaudhuri, G. Hockney and J. Lykken, \NPB{469}{1996}{357};\\
            J. Giedt, hep-th/0108244.

\bibitem{rffm}
I. Antoniadis, J. Ellis, J. Hagelin and D.V. Nanopoulos,
                                                \PLB{231}{1989}{65};\\
\AEF, D.V. Nanopoulos and K. Yuan, \NPB{335}{1990}{347};\\
I. Antoniadis, G.K. Leontaris, and J. Rizos, \PLB{245}{1990}{161};\\
\AEF, \PLB{278}{1992}{131}; \NPB{387}{1992}{239};\\
                G.B. Cleaver \etal, \PLB{455}{1999}{135};
                                    \NPB{620}{2002}{259};
                                    \PRD{63}{2001}{066001};
                                    hep-ph/0106060.

\bibitem{fccp}
X.G. Wen and E. Witten, \NPB{261}{1985}{651};\\
A. Schellekens, \PLB{237}{1990}{363};\\
J. Ellis, J.L. Lopez and D.V. Nanopoulos, \PLB{245}{1990}{375};
                                        \PLB{247}{1990}{375};\\ 
S. Chang, C. Corian\`{o} and A.E. Faraggi, \NPB{477}{1996}{257};\\
K.~Benakli, J.~Ellis, and D.V.~Nanopoulos, \PRD{59}{1999}{047301};\\
C. Corian\`{o}, A.E. Faraggi and M. Pl\"umacher, \NPB{614}{2001}{233}.

\bibitem{hw}
P. Ho\v rava and E. Witten, {\it Nucl. Phys.} {\bf B460} (1996) 506;
{\it Nucl. Phys.} {\bf B475} (1996) 94; for a review see \cite{mreviews}.

\bibitem{daow}
R. Donagi, A.  Lukas, B. Ovrut and D. Waldram,
                        {\it JHEP} {\bf 06} (1999) 034;\\
R. Donagi, B. Ovrut and D. Waldram, {\it JHEP} {\bf 11} (1999) 030;\\
E. Buchbinder, R. Donagi and B. Ovrut, {\tt hep-th/0202084}.


\bibitem{dlow}
R. Donagi, B. Ovrut, T. Pantev and D. Waldram, {\tt hep-th/9912208};
{\it Class. Quant. Grav.} {\bf 17} (2000) 1049.

\bibitem{flipped} S.M. Barr, \PLB{112}{1982}{219};\\
                  J.P. Derendinger, J.E. Kim and D.V. Nanopoulos,
                             \PLB{139}{1984}{170};\\
                  I. Antoniadis, J.R. Ellis, J.S. Hagelin and D.V. Nanopoulos,
                             \PLB{194}{1987}{231}.
\bibitem{vafa}
A. Iqbal, A. Neitzke and C. Vafa, {\tt hep-th/0111068}.

\bibitem{fmw}
R. Friedman, J. Morgan and E. Witten, {\it Comm. Math. Phys.} {\bf 187}
(1997) 679.

\bibitem{asian}
R. Donagi, {\it Asian J. Math.} {\bf 1} (1997) 214.

\bibitem{bersha}
M. Bershadsky {\it et al.}, {\it Nucl. Phys.} {\bf B505} (1997) 165. 

\bibitem{mova}
D. Morrison and C. Vafa,
{\it Nucl. Phys.} {\bf B473} (1996) 74;
{\it Nucl. Phys.} {\bf B476} (1996) 437.

\bibitem{bruzzo}
C. Bartocci, U. Bruzzo, D. Hern\'andez Ruip\'erez and J. Mu\~noz Porras,
{\it Comm. Math. Phys.} 195 (1998) 79.

\bibitem{curio}
G. Curio, {\it Phys. Lett.} {\bf B435} (1998) 39.

\bibitem{salamanca}
B. Andreas, G. Curio, D. Hern\'andez Ruip\'erez, S.-T. Yau,
{\tt math.AG/0012196}; {\it JHEP} {\bf 03} (2001) 020.

\bibitem{andreas}
B. Andreas, {\it JHEP} {\bf 01} (1999) 011.


\bibitem{bmr}
P. Berglund and P. Mayr, {\it Adv. Theor. Math. Phys.} {\bf 2} (1999) 1307;
{\it JHEP} {\bf 12} (1999) 009; \\
G. Rajesh, {\tt hep-th/9811240}.

\bibitem{calgsurf}
W. Barth, C. Peters and A. Van de Ven, {\it Compact Complex Surfaces},
Springer, Berlin (1984).

\bibitem{fff} {I. Antoniadis, C. Bachas, and C. Kounnas,
                                                \NPB{289}{1987}{87};\\
               H. Kawai, D.C. Lewellen, and S.H.-H. Tye, \NPB{288}{1987}{1}.}

\bibitem{nahe} \AEF~and D.V. Nanopoulos, \PRD{48}{1993}{3288}.

\bibitem{borceaviosin} C. Voisin, in
                  {\it Journ\'ees de G\'eom\'etrie Alg\'ebrique
                    d'Orsay} (Orsay, 1992), Ast\'erisque {\bf 218} (1993)
273;\\
                      C. Borcea, in {\it Essays on Mirror Manifolds}, Vol
            2,(B. Greene and S.-T. Yau, eds.), International Press, Cambridge,
            1997, p. 717;\\
                        V. Nikulin, in
                   {\it Proceedings of the International
                         Congress of Mathematicians} (Berkeley, 1986),
                         p. 654.
\bibitem{mukhi}
R. Gopakumar and S. Mukhi, {\it Nucl. Phys.} {\bf B479} (1996) 260.

\bibitem{campbell} B.A. Campbell \etal, \PLB{198}{1987}{200}.

\bibitem{fm} J.L. Lopez and D.V. Nanopoulos, \NPB{338}{1990}{73};
                \PLB{251}{1990}{73}; \PLB{268}{1991}{359};\\ 
             \AEF, \NPB{403}{1993}{101}; \NPB{407}{1993}{57};\\
             \AEF~and E. Halyo, \NPB{416}{1994}{63};\\
             J. Giedt, \NPB{595}{2001}{3}.

\bibitem{inprep} \AEF, R. Garavuso and J.M. Isidro, paper in preparation.

\bibitem{foc} \AEF, \PLB{326}{1994}{62}.

\bibitem{efn3}
J. Ellis \etal, \PLB{419}{1998}{123};
\PLB{433}{1998}{269}; \IJMP{15}{2002}{1345}.



\end{thebibliography}
\end{document}